# Sub-60 mV/decade switching in a metal-insulator-metal-insulator-semiconductor transistor without ferroelectric component

*Peng Wu[1,2*], Joerg Appenzeller[1,2*]*

[1]Birck Nanotechnology Center, Purdue University, West Lafayette, Indiana 47907, USA

[2]Department of Electrical and Computer Engineering, Purdue University, West Lafayette, Indiana 47907, USA

[*]E-mail: wu936@purdue.edu, appenzeller@purdue.edu

**Abstract:**

Negative capacitance field-effect transistors (NC-FETs) have attracted wide interest as promising candidates for steep-slope devices, and sub-60 mV/decade switching has been demonstrated in NC-FETs with various device structures and material systems. However, the detailed mechanisms of the observed steep-slope switching in some of these experiments are under intense debate. Here we show that sub-60 mV/decade switching can be observed in a $WS_2$ transistor with a metal-insulator-metal-insulator-semiconductor (MIMIS) structure – without any ferroelectric component. This structure resembles an NC-FET with internal gate, except that the ferroelectric layer is replaced by a leaky dielectric layer. Through simulations of the charging dynamics during the device characterization using an RC network model, we show that the observed steep-slope switching in our "ferroelectric-free" transistors can be attributed to the internal gate voltage response to the chosen varying gate voltage scan rates. We further show that a constant gate voltage scan rate can also lead to transient sub-60 mV/decade switching in an MIMIS structure with voltage dependent internal gate capacitance. Our results indicate that the observation of sub-60 mV/decade switching alone is not sufficient evidence for the successful demonstration of a true steep-slope switching device and that experimentalists need to critically assess their measurement setups to avoid measurement-related artefacts.

Negative capacitance field-effect transistors (NC-FETs) have been proposed as devices that could break the 60 mV/decade subthreshold swing (SS) limit of metal-oxide-semiconductor field-effect transistors (MOSFETs) at room temperature and achieve low-power switching[1–4]. The proposed NC-FET structure includes a ferroelectric (FE) capacitor in series with a normal dielectric (DE) capacitor to stabilize the negative capacitance region in the ferroelectric material, leading to an internal voltage amplification and sub-60 mV/decade switching. Since then, sub-60 mV/decade operations have been reported in NC-FETs from various material systems and device structures[5–10]. However, the origin of the observed steep-slope switching is still under intense debate. Different from the original proposal, alternative mechanisms for the observed steep slope switching have been proposed and the validity of a stabilizing negative capacitance (NC) region of FE materials has been challenged[11–13]. For example, some papers relate the observed steep switching to transient effects such as internal potential jumps[14,15] or differential voltage amplification caused by polarization switching of FE material[16–22], instead of a quasi-static NC effect represented as the "S"-shaped curve in the polarization–electric field ($P$–$E$) relation of FE materials[1,23,24]. Other explanations include domain nucleation and growth in multi-domain FE materials[15,25,26]. Yet due to the complexity of both the ferroelectric physics and the experimental systems, there has not been a single consolidated theory that can convincingly explain all the experimental findings. Therefore, a model system exhibiting similar device characteristics yet with simplified physics is highly sought after.

In this work, we experimentally demonstrate a two-dimensional tungsten disulfide ($WS_2$) transistor in a metal-insulator-metal-insulator-semiconductor (MIMIS) device structure, in which sub-60 mV/decade switching is observed. The device structure is similar to an NC-FET with internal metal gate, or equivalently a metal-ferroelectric-metal-insulator-semiconductor (MFMIS) structure (Fig. 1a), except that the ferroelectric (FE) layer is replaced by a leaky dielectric (DE) layer (Fig. 1b). In previous publications, negative capacitance FETs with leaky FE layers have been studied[27,28]. However, to the best of our knowledge, there have been no discussions on the impact of leaky insulators in devices built with only DE layers and no FE layers. Also it has been reported that a lossy dielectric can be misinterpreted as ferroelectric due to measurement artefacts[29], yet there have been no discussions on the implications of such artefacts for electrical devices.

In the MIMIS device, we show that several key phenomena observed in NC-FETs can be reproduced (Figs. 1c & 1d): (i) sub-60 mV/decade switching, (ii) anti-clockwise hysteresis, and (iii) internal voltage amplification. By replacing the FE layer with a DE layer, we are able to avoid the controversial discussion about polarization switching in the ferroelectric and can focus entirely on the charging dynamics inside the system, which is built with well-understood basic circuit components such as resistors (R) and capacitors (C). Using a simple RC network model, we are able to simulate the charging dynamics in the MIMIS device. Distinct device features i) through iii) from above can be readily explained as a result of the varying gate voltage scan rates due to auto-ranging of the current measurement, which could be a common issue in the characterization of steep-slope transistors. We further show that a sub-60 mV/decade switching can still be observed with a constant gate scan rate if the internal gate capacitance is voltage-dependent, highlighting the universality of the effect.

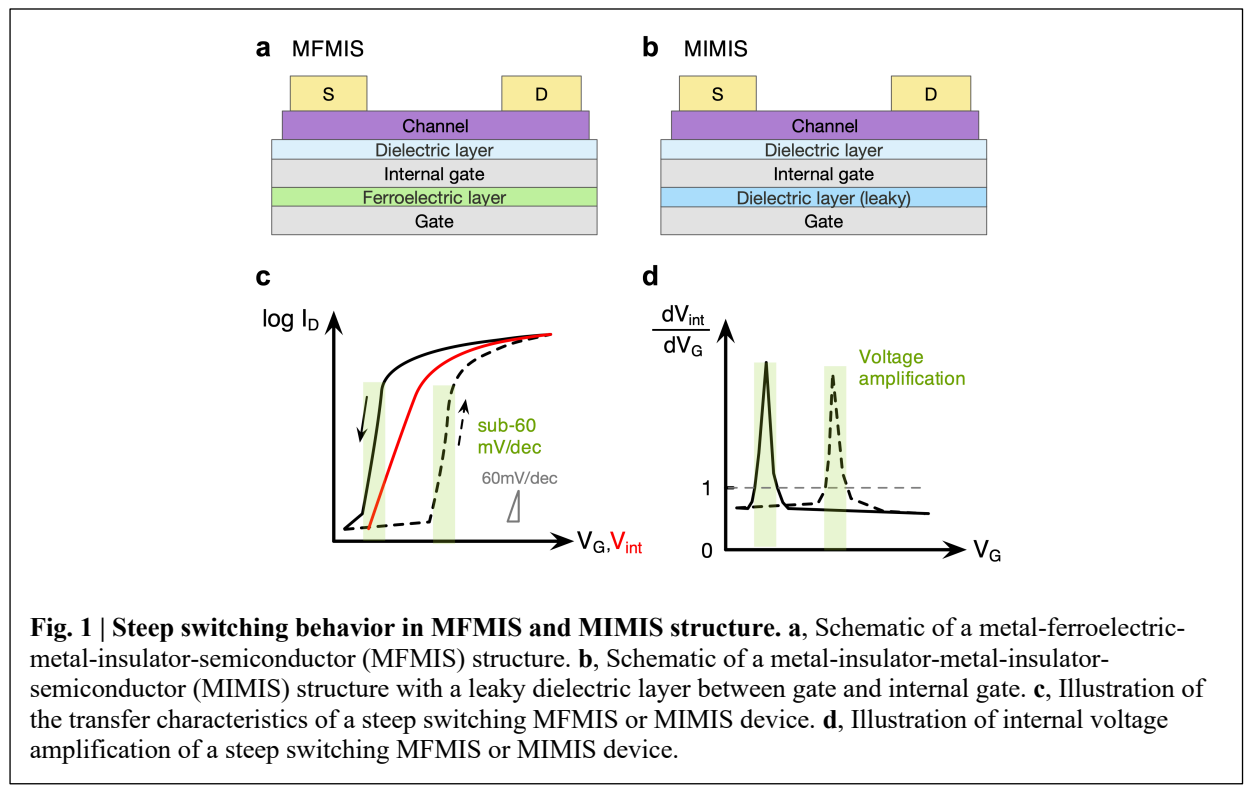

**Fig. 1 | Steep switching behavior in MFMIS and MIMIS structure. a**, Schematic of a metal-ferroelectric-metal-insulator-semiconductor (MFMIS) structure. **b**, Schematic of a metal-insulator-metal-insulator-semiconductor (MIMIS) structure with a leaky dielectric layer between gate and internal gate. **c**, Illustration of the transfer characteristics of a steep switching MFMIS or MIMIS device. **d**, Illustration of internal voltage amplification of a steep switching MFMIS or MIMIS device.

## Results

**Experimental demonstration of sub-60 mV/decade switching in an MIMIS device.** Fig. 2a shows a schematic of the MIMIS device. The internal FET is composed of an 8-nm-thick $WS_2$ channel with Ni source/drain metal contacts on top of a local bottom gate structure with 7-nm-thick $HfO_2$ gate dielectric grown by atomic layer deposition (ALD) and Ti gate metal as "Internal gate" electrode, which is separated from the "Gate" electrode by a low-quality, leaky $AlO_x$ dielectric film grown by natural oxidation of a thin Al layer (see Methods). The leakage current through the leaky $AlO_x$ dielectric film in a capacitor with an area of ~1 μm$^2$ is on the order of 1~100 pA at ±0.5 V (Fig. 2b).

Fig. 2c shows the $I_D - V_{int}$ transfer characteristics of the internal $WS_2$ FET. All device characterization in this study was performed using an Agilent 4156C semiconductor parameter analyzer in a LakeShore FWPX probe station at a vacuum level below 10$^{-5}$ Torr at room temperature. Due to the ultra-thin body of the $WS_2$ channel and high quality of the ALD-grown $HfO_2$ dielectric[30], an SS-value of 61 mV/decade is achieved in the $WS_2$ internal FET, close to the 60 mV/decade room-temperature limit. Also note that the hysteresis in the $I_D - V_{int}$ characteristics of the $WS_2$ internal FET is negligible (inset of Fig. 2c) due to the high quality of the $HfO_2$ dielectric and low defect density at the semiconductor-oxide interface.

Next, we characterize the $I_D - V_G$ transfer characteristics of the $WS_2$ MIMIS device. As shown in Fig. 2d, an anti-clockwise hysteresis is observed in the $I_D - V_G$ characteristics, and sub-60 mV/decade switching behaviors are observed in both sweep directions, indicated by the green and orange shaded regions, respectively. The observed characteristics resemble previous reported two-dimensional (2D) NC-FETs with MFMIS device structures[6–8], even though there is no ferroelectric component in our MIMIS device. Since the $I_D - V_{int}$ characteristics of the internal $WS_2$ FET are essentially hysteresis-free, we can derive a one-to-one mapping between $I_D$ and $V_{int}$ at a given drain voltage and extract $V_{int}$ (Fig. 2e) and the internal voltage gain $\delta V_{int}/\delta V_G$ (Fig. 2f) during the $I_D - V_G$ scan. One can see that internal voltage amplifications ($\delta V_{int}/\delta V_G > 1$) can be observed in both sweeping directions in the $I_D - V_G$ scan. Note that although direct measurement methods of the internal gate voltage $V_{int}$ during the $I_D - V_G$ scan exist, previous studies[14,15] have pointed out that a voltage measurement system with input impedance much larger than the leakage resistances of the capacitors is required to accurately measure these voltages, which is not available in our measurement system. Therefore, we have adopted the strategy of indirectly extracting $V_{int}$ from the measured $I_D$.

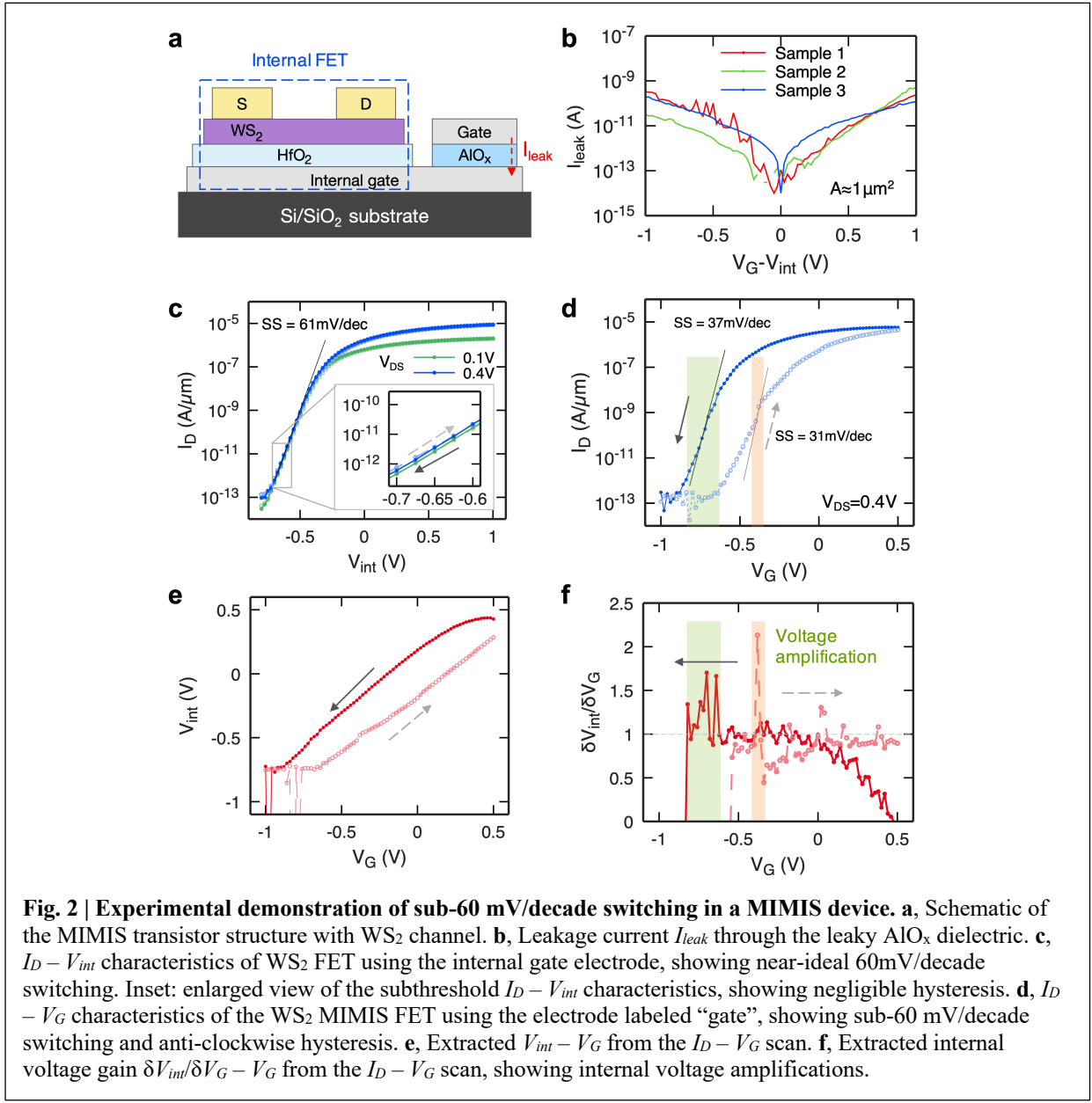

**Fig. 2 | Experimental demonstration of sub-60 mV/decade switching in a MIMIS device. a**, Schematic of the MIMIS transistor structure with WS$_2$ channel. **b**, Leakage current $I_{leak}$ through the leaky AlO$_x$ dielectric. **c**, $I_D - V_{int}$ characteristics of WS$_2$ FET using the internal gate electrode, showing near-ideal 60mV/decade switching. Inset: enlarged view of the subthreshold $I_D - V_{int}$ characteristics, showing negligible hysteresis. **d**, $I_D - V_G$ characteristics of the WS$_2$ MIMIS FET using the electrode labeled "gate", showing sub-60 mV/decade switching and anti-clockwise hysteresis. **e**, Extracted $V_{int} - V_G$ from the $I_D - V_G$ scan. **f**, Extracted internal voltage gain $\delta V_{int}/\delta V_G - V_G$ from the $I_D - V_G$ scan, showing internal voltage amplifications.

**Modelling of the charging dynamics in the MIMIS device.** To understand the mechanism behind the observed sub-60 mV/decade switching behaviour and internal voltage amplification in the WS$_2$ MIMIS device, we have performed circuit simulations of the charging dynamics during the $I_D$ – $V_G$ scan based on an equivalent circuit model shown in Fig. 3a. In the equivalent circuit, we are considering the quantum capacitance of the WS$_2$ channel $C_q$, the oxide capacitance of HfO$_2$ dielectric $C_{ox}$, the parasitic capacitance of the internal gate to the substrate $C_p$, and the capacitance of the AlO$_x$ dielectric between gate and the internal gate $C_{int-G}$. The leakage current in the AlO$_x$ capacitor is modelled as a linear resistor $R$ to simplify the discussion, although the leakage current is non-linear with voltage. Since the parasitic capacitance of the internal gate to substrate $C_p$ is much larger than $C_{ox}$ and $C_q$ (see Supplementary Section 1), it is reasonable to consider only $C_p$ and ignore $C_{ox}$ and $C_q$ in the simulation of the $V_{int}$ response. The parameters used in the simulation are (Supplementary Section 1):

$$C_p = 9.59 \times 10^{-13} \text{ F}, C_{int-G} = 7.75 \times 10^{-15} \text{ F}, R = 6 \times 10^{11} \text{ } \Omega$$

Next, we calculate the $V_{int}$ response to a given $V_G$ scan. From Kirchhoff's Current Law (KCL) at the $V_{int}$ node, we have:

$$I_{\text{leak}} = \frac{V_G - V_{\text{int}}}{R} = \frac{d}{dt}[C_p V_{\text{int}} - C_{\text{int}-G}(V_G - V_{\text{int}})] \quad (1)$$

$$\frac{dV_{\text{int}}}{dt} = \frac{V_G - V_{\text{int}}}{R(C_p + C_{\text{int}-G})} + \frac{C_{\text{int}-G}}{C_p + C_{\text{int}-G}} \frac{dV_G}{dt} \quad (2)$$

Let $dV_{int}/dt = dV_G/dt$, we have the steady state response:

$$V_G - V_{\text{int}} = RC_p \frac{dV_G}{dt} \quad (3)$$

As indicated by Eq. (3), the difference between $V_{int}$ and $V_G$ at steady state is proportional to the RC time constant (which is 0.58 s for the parameters chosen here) and the $V_G$ sweep rate (d$V_G$/d$t$), which agrees with the simulated $V_{int}$ responses to different $V_G$ scan rates in Fig. 3b. Now if we think about a *varying the $V_G$ sweep rate* ("fast" to "slow", Fig. 3c), there is a transition period (green shaded area in Fig. 3c) during which the difference between $V_{int}$ and $V_G$ changes from one steady state to another, and in this transition period we have $|dV_{int}/dt| > |dV_G/dt|$, which indicates a differential internal voltage amplification ($\delta V_{int}/\delta V_G > 1$).

From the above analysis, it is clear that in the RC circuit model, differential internal voltage amplification occurs when the gate sweep rate is changed from "fast" to "slow". Next, we identify such events in the experimental $I_D - V_G$ measurement of the WS$_2$ MIMIS device. When measuring the transfer characteristics of a transistor, since the current range may span more than 7 orders of magnitude (as shown in Figs. 2c & 2d), auto ranging is typically adopted by the semiconductor parameter analyzer[31], and it is common to use slower scan rates for smaller current ranges to allow for longer integration times to reduce errors. On the other hand, faster scan rates are used for larger current ranges to increase measurement speed. In our measurement setup (see Methods), a "fast" scan rate of 0.36 V/s is adopted when the current level is above 1 nA (~0.6 nA/μm in the $I_D - V_G$ curve, normalized by channel width $W$=1.7 μm), and a "slow" scan rate of 0.208 V/s is adopted when the current level is below 1 nA (Supplementary Section 2 and Supplementary Video 1).

Based on the findings, we reconstruct the $V_G$ input as a function of time[†] during the $I_D - V_G$ scan of the WS$_2$ MIMIS device at $V_{DS}$ = 0.4 V and simulate the $V_{int}$ response based on Eq. (2), as shown in Fig. 3d. An initial condition $V_{int}(t = 0) = 0.43$ V is determined from the first datapoint in the measured $V_{int} - V_G$ shown in Fig. 2e. We monitor $V_{int}$ and the respective $I_D$ in the simulation and change the $V_G$ sweep rate correspondingly when the current level reaches values above or below 1 nA to mimic the experimental setup. In the forward scan direction (0.5 V to -1 V) of the $V_G$ sweep, we observe a differential internal voltage amplification when the $V_G$ sweep rate changes from "fast" to "slow", indicated by the green shaded area in Fig. 3e, which is in accordance with our previous analysis. In the backward scan direction

---

[†] In the real measurement, the $V_G$ input should be a staircase function of time $t$, instead of a linear function used in the simulation. Yet due to the small step size (0.02 V), the difference is negligible.

(-1 V to 0.5 V), however, the scan rate change from "slow" to "fast" and should not result in the observed sub-60 mV/decade switching and differential voltage amplification shown in Figs. 2d & 2f. The reason for the observed phenomenon, i.e. again voltage amplification in the experiment, is that an overhead time $t_{overhead} \approx 0.22$ s is associated with the change of current measurement range[‡], as indicated by the orange shaded area in Fig. 3f. During the overhead time, $V_G$ increases *only* by the scan step size $V_{step} = 0.02$ V, resulting in an equivalent scan rate of 0.091 V/s. Since $dV_{int}/dt$ remains almost unchanged [Eq. (2)] while $dV_G/dt$ slows down significantly during the overhead time, a transient differential internal voltage amplification $|dV_{int}/dt| > |dV_G/dt|$ occurs. Figs. 3g & 3h show the simulated internal gate voltage $V_{int}$ and drain current $I_D$ as a function of $V_G$, which shows good agreement with the experimental data. The experimentally observed characteristics, such as anti-clockwise hysteresis and sub-60 mV/decade switching in both sweeping directions, are reproduced in the simulation. The transition points of scan rates are indicated by the green and orange lines in the figures. Fig. 3i shows the simulated internal voltage gain $\delta V_{int}/\delta V_G$ as a function of $V_G$, which shows good agreement with the experimental data not only qualitatively, but also quantitatively in terms of amplitudes of the internal voltage amplifications and the corresponding $V_G$ positions. This clearly shows that our simple model captures the essential characteristics of the charging dynamics of the $I_D - V_G$ measurement. Note that from Eq. (3), we can see that the voltage difference between $V_{int}$ and $V_G$ in the steady state is given by $RC_P \cdot (dV_G/dt)$, which is equal to 0.21 V and 0.12 V for "slow" and "fast" scan rates, respectively, and both are smaller than the voltage scan window of 1.5 V. If the RC time constant is too small, $V_{int}$ would follow $V_G$ completely; if the RC time constant is too large, $V_{int}$ would not reach steady state within the voltage scan window. Therefore, the RC time constant needs to be comparable to the timescale of the $V_G$ scan in order to see the described effects.

---

[‡] Although not specified in the manual of the semiconductor parameter analyzer[31], our measurement results indicate that the overhead time only occurs at the transition from a smaller current measurement range into a larger measurement range, and not vice versa.

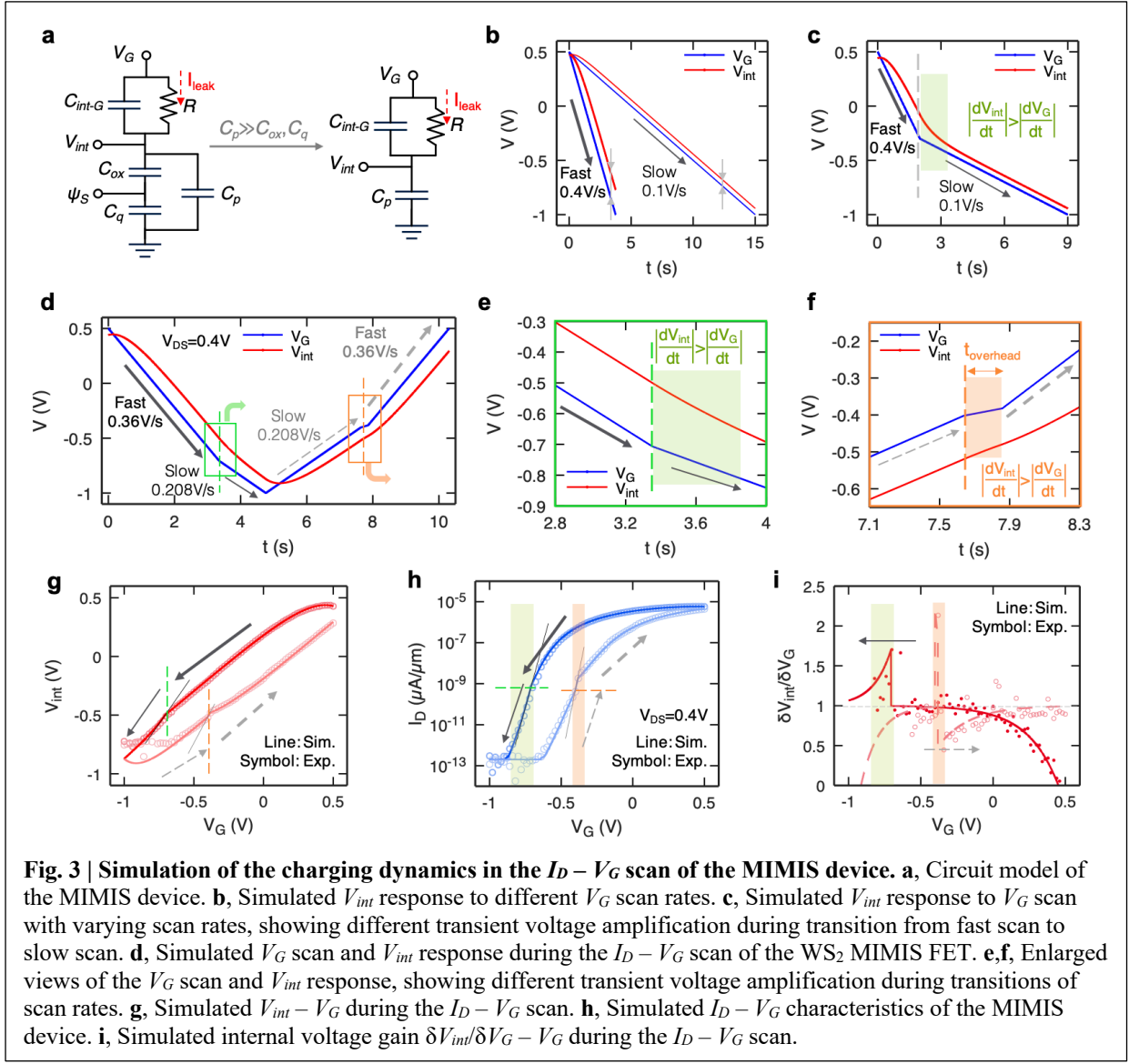

**Fig. 3 | Simulation of the charging dynamics in the $I_D - V_G$ scan of the MIMIS device. a**, Circuit model of the MIMIS device. **b**, Simulated $V_{int}$ response to different $V_G$ scan rates. **c**, Simulated $V_{int}$ response to $V_G$ scan with varying scan rates, showing different transient voltage amplification during transition from fast scan to slow scan. **d**, Simulated $V_G$ scan and $V_{int}$ response during the $I_D - V_G$ scan of the WS$_2$ MIMIS FET. **e,f**, Enlarged views of the $V_G$ scan and $V_{int}$ response, showing different transient voltage amplification during transitions of scan rates. **g**, Simulated $V_{int} - V_G$ during the $I_D - V_G$ scan. **h**, Simulated $I_D - V_G$ characteristics of the MIMIS device. **i**, Simulated internal voltage gain $\delta V_{int}/\delta V_G - V_G$ during the $I_D - V_G$ scan.

**Simulation of sub-60 mV/decade switching with constant $V_G$ scan rate.** We have simulated the characteristics of the WS$_2$ MIMIS device with varying $V_G$ scan rates and have shown that transient sub-60 mV/decade switching can be observed due to internal voltage amplifications when the scan rate is changing. However, one may argue that while varying scan rates are common when characterizing transistors, a constant scan rate is achievable with proper configuration of measurement setups and thus the effects discussed in the previous section can be avoided. To show that such effects are rather universal and do not just apply to a particular scenario, we simulate the case of a constant $V_G$ scan rate and show that sub-60 mV/decade switching can still be observed.

In the previous section, the circuit under discussion had a constant RC time constant due to a large, constant parasitic capacitance $C_p$, and in combination with a time-varying $V_G$ scan rate, sub-60 mV/decade switching can be observed. One can envision that with a constant $V_G$ scan rate, yet in combination with a *varying* RC time "constant", similar effects could also be achieved. It is well-known that the gate capacitance of a MOSFET is voltage-dependent[32], and for the 2D WS$_2$ transistor in this study, $C_{MOS}$ can be represented by the series of oxide capacitance $C_{ox}$ and quantum capacitance[33,34] $C_q$, which can be expressed by:

$$C_q = q^2 \text{DOS}_{2D} f(E_C) A_{\text{device}} = q^2 \frac{g_C m_e^*}{\pi \hbar^2} \frac{1}{1+\exp[(E_C - E_F)/k_B T]} A_{\text{device}} \qquad (4)$$

where $g_C$ is conduction band valley degeneracy, $m_e^*$ is electron effective mass. From Eq. (4), we can see that $C_q$ is dependent on the relative position of conduction band edge $E_C$ and Fermi level $E_F$, which can be tuned with gate voltage, and by removing the parasitic capacitance from the circuit, we can achieve a voltage-dependent internal gate capacitance $C_{MOS}$ for the internal FET (Fig. 4a). Fig. 4b shows the simulated $C_{ox}$, $C_q$ and $C_{MOS}$ as a function of $V_{int}$ (see Supplementary Section 3). One can see that when $V_{int}$ is negative and the internal WS$_2$ FET is in its off-state, $C_q$ is nearly zero as the Fermi level $E_F$ in WS$_2$ is in the bandgap, and thus $C_{MOS}$ is almost zero; when $V_{int}$ increases and the internal WS$_2$ FET is turned on, $C_q$ quickly increases and exceeds $C_{ox}$ as the Fermi level $E_F$ approaches the conduction band edge $E_C$, and thus $C_{MOS}$ approaches $C_{ox}$.

Next, we simulate the $V_{int}$ response with a constant $V_G$ scan rate and voltage-dependent $C_{MOS}$, for which Eq. (1) and Eq. (2) are modified as follows:

$$I_{\text{leak}} = \frac{V_G - V_{\text{int}}}{R} = \frac{d}{dt}[Q_{\text{ch}}(V_{\text{int}}) - C_{\text{int}-G}(V_G - V_{\text{int}})] = \frac{\partial Q_{\text{ch}}}{\partial V_{\text{int}}}\frac{dV_{\text{int}}}{dt} - C_{\text{int}-G}\frac{d(V_G - V_{\text{int}})}{dt} \quad (5)$$

$$= [C_{\text{MOS}}(V_{\text{int}}) + C_{\text{int}-G}]\frac{dV_{\text{int}}}{dt} - C_{\text{int}-G}\frac{dV_G}{dt}$$

$$\frac{dV_{\text{int}}}{dt} = \frac{V_G - V_{\text{int}}}{R[C_{\text{MOS}}(V_{\text{int}}) + C_{\text{int}-G}]} + \frac{C_{\text{int}-G}}{C_{\text{MOS}}(V_{\text{int}}) + C_{\text{int}-G}}\frac{dV_G}{dt} \quad (6)$$

From Eq. (6), we simulate the $V_{int}$ response to the $V_G$ scan with a constant scan rate of 1 V/s, as shown in Fig. 4c. We have also changed the parameter $R = 2\times10^{13}$ Ω to compensate for the removal of a large capacitance $C_p$ to keep the RC time constant approximately the same ($RC_{ox} = 0.32$ s). A differential internal voltage amplification $|dV_{int}/dt| > |dV_G/dt|$ (green shaded area) is observed in the forward scan direction (0.5 V to -1 V), but not in the backward scan direction (-1 V to 0.5 V). Fig. 4d shows the simulated $V_{int}$ as a function of $V_G$. Assuming the same $I_D - V_{int}$ as in Fig. 2c, $I_D - V_G$ characteristics can be simulated (Fig. 4e), in which a sub-60 mV/decade switching is observed in the forward scan direction. Fig. 4f shows the simulated internal voltage gain $\delta V_{int}/\delta V_G$ as a function of $V_G$, in which the differential internal voltage amplification is clearly illustrated by the green shaded area.

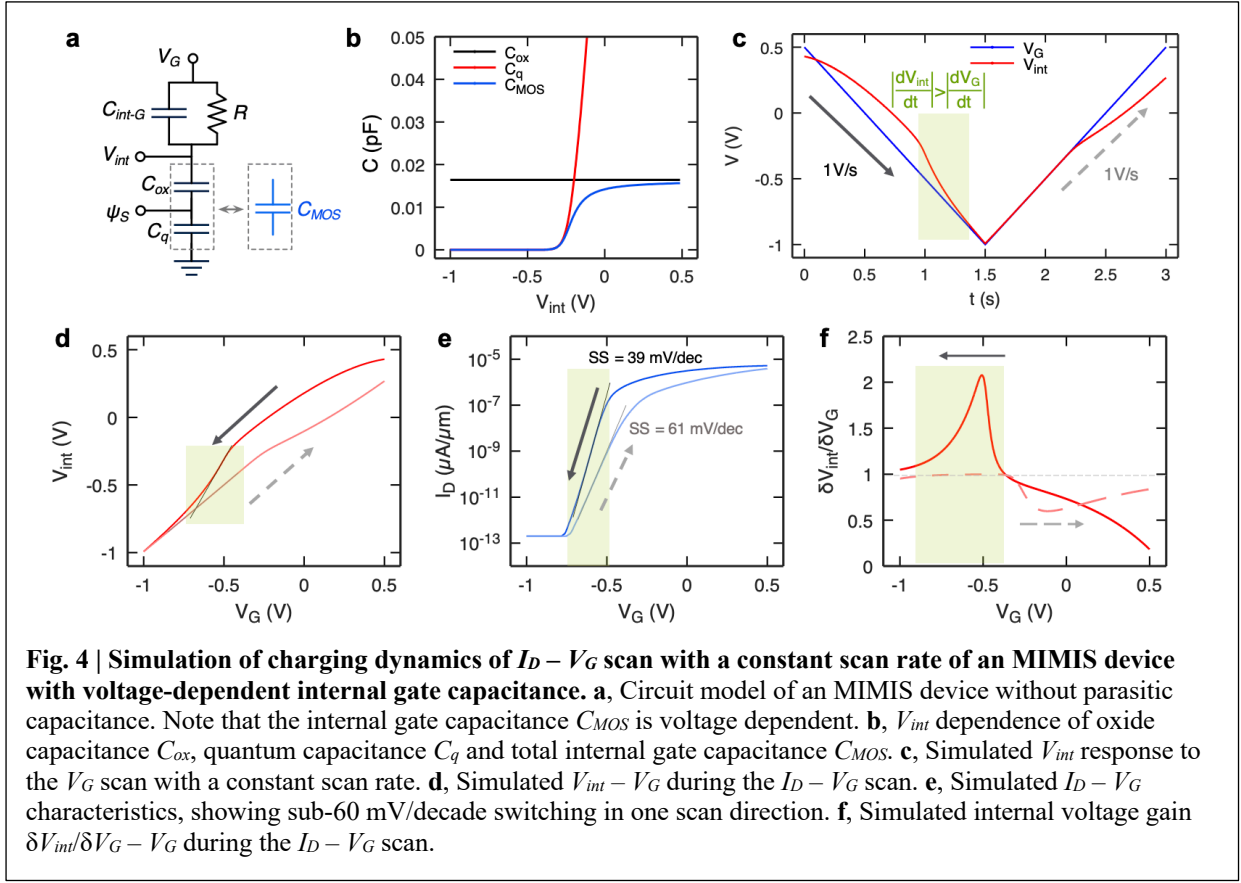

**Fig. 4 | Simulation of charging dynamics of $I_D - V_G$ scan with a constant scan rate of an MIMIS device with voltage-dependent internal gate capacitance. a**, Circuit model of an MIMIS device without parasitic capacitance. Note that the internal gate capacitance $C_{MOS}$ is voltage dependent. **b**, $V_{int}$ dependence of oxide capacitance $C_{ox}$, quantum capacitance $C_q$ and total internal gate capacitance $C_{MOS}$. **c**, Simulated $V_{int}$ response to the $V_G$ scan with a constant scan rate. **d**, Simulated $V_{int} - V_G$ during the $I_D - V_G$ scan. **e**, Simulated $I_D - V_G$ characteristics, showing sub-60 mV/decade switching in one scan direction. **f**, Simulated internal voltage gain $\delta V_{int}/\delta V_G - V_G$ during the $I_D - V_G$ scan.

## Discussion

In previous demonstrations of NC-FETs, observing sub-60 mV/decade switching has been deemed as a mark of success for steep-slope switching devices. However, the results in this work show that by putting a leaky capacitor in series with the gate of an FET, which is clearly *not* a solution to steep-slope switching devices, sub-60 mV/decade switching can also be observed. Such effects originate from transient voltage amplifications over a small voltage window ($\delta V_{int}/\delta V_G > 1$), instead of persistent voltage amplifications over the entire voltage scan range ($\Delta V_{int}/\Delta V_G > 1$) that is required for a real demonstration of low-voltage switching. The transient voltage amplification may occur during transitions between current measurement ranges, where changes in the voltage scan rates are typical, and even for a constant scan rate, an abrupt change of internal gate capacitance near threshold voltage $V_{th}$ may lead to similar effects. Since the effects are caused by the charging dynamics in a non-ideal measurement setup, which is likely to be utilized in the characterization of NC-FETs by many, it is imperative to reassess the measurement setups in the previous reported sub-60 mV/decade switching NC-FETs with internal metal gates, to rule out such measurement-related artefacts and verify the existence of real voltage amplification predicted by the quasi-static NC theory.

**Methods**

Device fabrication

Internal gate electrodes (2 nm Ti) were deposited using e-beam evaporation onto a silicon substrate with 90 nm of silicon dioxide on top and patterned using standard e-beam lithography and lift-off process. 7 nm of $HfO_2$ was deposited using atomic layer deposition (ALD) on top of the internal gate metal as dielectric for the internal $WS_2$ FET. The $HfO_2$ film was grown using a low-temperature (90 °C) ALD process and patterned using e-beam lithography and lift-off to expose part of the internal gate metal. Few-layer $WS_2$ flakes were exfoliated onto the local bottom gate structure from bulk crystals (purchased from HQGraphene). Ni (40 nm) electrodes were deposited and patterned by e-beam evaporation and e-beam lithography as source/drain contacts for the $WS_2$ FET. A thin aluminium film was deposited on the exposed internal gate metal using e-beam evaporation and oxidized in air for 15 minutes to form the leaky $AlO_x$ dielectric. To make sure the Al is fully oxidized, the evaporation-oxidization was done in two steps (evaporation-oxidation-evaporation-oxidation) with 2 nm Al in each deposition, instead of 4 nm in a single deposition step. Finally, the gate metal (40 nm Ti) was deposited using e-beam evaporation and patterned using e-beam lithography and lift-off process.

Device characterization

The electrical characterization was performed in a LakeShore FWPX Probe Station at a vacuum level below $10^{-5}$ Torr using an Agilent 4156C Parameter Analyzer. All measurements were performed at room temperature. The sweep measurement in the parameter analyzer has the following settings:

Primary sweep (VAR1): VG, start voltage: 0.5 V, end voltage: -1 V, step voltage: 0.02 V, double sweep: enabled, compliance: 1e-7 A

Subordinate sweep (VAR2): VD, start voltage: 0.1 V, step voltage: 0.3 V, no. of steps: 2, compliance: 1e-1 A

Measurement ranging mode: auto ranging, integration time: SHORT, hold time: 0.5 s, delay time: 0 s

**Data availability:**

The data that support the findings within this paper are available from the corresponding authors on reasonable request.

**Code availability:**

The computer code used in this study is available from the corresponding authors upon reasonable request.

**Acknowledgements:**

This work was supported in part by the Semiconductor Research Corporation (SRC) at the NEW LIMITS Center and National Institute of Standards and Technology (NIST) through award number 70NANB17H041

**Author contributions:**

P.W. and J.A. conceived the idea and designed the experiments. P.W. performed the device fabrication, characterization and the simulation. Both authors wrote the manuscript.

**Competing interests:**

The authors declare no competing interests.

**Additional information:**

Correspondence and requests for materials should be addressed to P.W. and J.A.

**Supplementary Information for**

**Sub-60 mV/decade switching in a metal-insulator-metal-insulator-semiconductor transistor without ferroelectric component**

*Peng Wu[1,2*], Joerg Appenzeller[1,2*]*

[1]Birck Nanotechnology Center, Purdue University, West Lafayette, Indiana 47907, USA

[2]Department of Electrical and Computer Engineering, Purdue University, West Lafayette, Indiana 47907, USA

[*]E-mail: wu936@purdue.edu, appenzeller@purdue.edu

# 1. Determination of capacitance and resistance components in the RC network model

Supplementary Fig. 1a shows a schematic of the capacitance and resistance components in the WS$_2$ MIMIS device. The overlapping area of internal gate and gate is $A_{int\text{-}G} = 1$ μm$^2$ and the area of the contact pad for the internal gate is $A_{pad} = 50$ μm × 50 μm = 2500 μm$^2$, respectively. The device area of the WS$_2$ FET is $A_{device} = 1$ μm$^2$. The capacitances can be calculated:

$$C_{\text{int}-G} = \frac{\varepsilon_{AlO_x}\varepsilon_0 A_{\text{int}-G}}{t_{AlO_x}} = \frac{7 \times 8.854 \times 10^{-12} \times 1 \times (10^{-6})^2}{8 \times 10^{-9}} = 7.75 \times 10^{-15}\,\text{F} \tag{1}$$

$$C_p = \frac{\varepsilon_{SiO_2}\varepsilon_0 A_{pad}}{t_{SiO_2}} = \frac{3.9 \times 8.854 \times 10^{-12} \times (50 \times 10^{-6})^2}{90 \times 10^{-9}} = 9.59 \times 10^{-13}\,\text{F} \tag{2}$$

$$C_{ox} = \frac{\varepsilon_{HfO_2}\varepsilon_0 A_{device}}{t_{HfO_2}} = \frac{13 \times 8.854 \times 10^{-12} \times (1 \times 10^{-6})^2}{7 \times 10^{-9}} = 1.64 \times 10^{-14}\,\text{F} \tag{3}$$

The calculation of $C_q$ can be found in Supplementary Section 3, yet since it's in series with $C_{ox}$, the total capacitance $C_{MOS}$ is smaller than $C_{ox}$. We can see that $C_{ox}$ is much smaller than $C_p$. Therefore, we can ignore $C_{MOS}$ when simulating the $V_{int}$ response.

Next, the resistance value for $R$ is determined. Supplementary Fig. 1b shows the leakage current $I_{leak}$ in the AlO$_x$ capacitor as a function of voltage difference $V_{diff} = V_G - V_{int}$, which clearly exhibits a non-linear behavior and different $R$ values can be extracted at different $V_{diff}$-values. In addition, due to inaccuracies and noises in the measurement, there is a significant uncertainty associated with the extracted $R$ value, especially when the current is small. Therefore, we have adopted an indirect method of extracting $R$ by fitting the measured $V_{int} - V_G$ curve, as shown in Supplementary Fig. 1c. We have derived in the main text that, in the steady state:

$$V_G - V_{\text{int}} = RC_p\frac{\text{d}V_G}{\text{d}t} \tag{4}$$

Therefore, different $R$ values lead to different $V_G$-$V_{int}$ values and thus different hysteresis voltage values in the bidirectional $V_G$ scan. As clearly shown in Supplementary Fig. 1c, $R = 600$ MΩ gives the best fit to the experimental data, while other values either underestimate or overestimate the hysteresis voltage. The good fitting also indicates that using a fixed value for *R* is a good approximation despite the non-linear $I_{leak} - V_{diff}$ characteristics. Supplementary Fig. 1d shows the simulated $V_{int}$-$V_G$ as a function of time using the parameter $R = 600$ MΩ. The voltage difference is below 0.21 V during the $V_G$ voltage scan, and from the $I_{leak} - V_{diff}$ characteristics in Supplementary Fig 1b, we can roughly estimate $R \approx 900$ MΩ at $V_G$-$V_{int}$ = -0.2 V, which is in acceptable agreement with the extracted $R = 600$ MΩ.

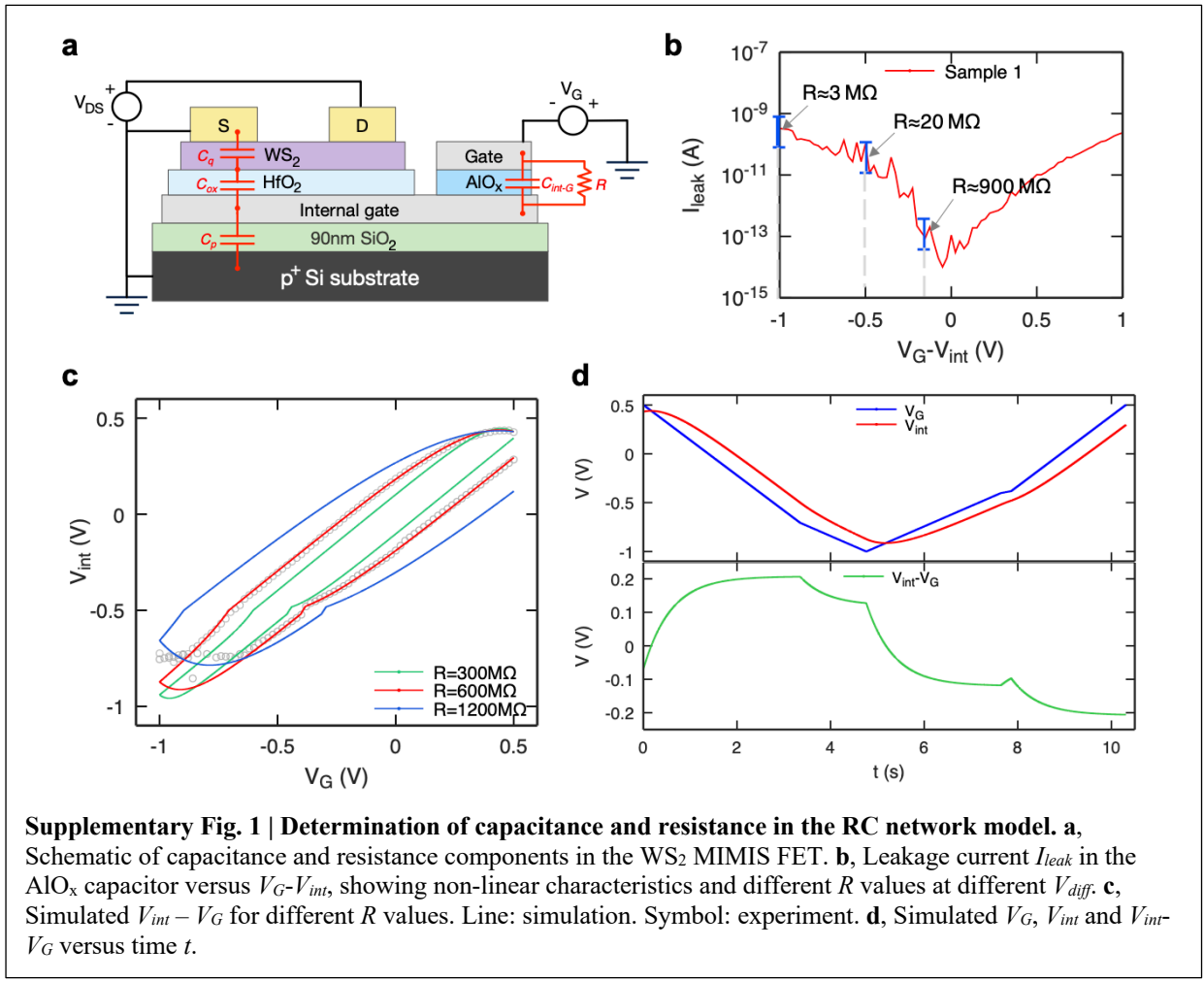

**Supplementary Fig. 1 | Determination of capacitance and resistance in the RC network model. a**, Schematic of capacitance and resistance components in the WS$_2$ MIMIS FET. **b**, Leakage current $I_{leak}$ in the AlO$_x$ capacitor versus $V_G$-$V_{int}$, showing non-linear characteristics and different *R* values at different $V_{diff}$. **c**, Simulated $V_{int} - V_G$ for different *R* values. Line: simulation. Symbol: experiment. **d**, Simulated $V_G$, $V_{int}$ and $V_{int}$-$V_G$ versus time *t*.

## 2. Extraction of scan rates from experiment

Supplementary Video 1 shows a video of a researcher performing an $I_D - V_G$ scan of the WS$_2$ MIMIS FET on the Agilent 4156C parameter analyzer (a single frame of the video is shown in Supplementary Fig. 2a as an example). The measured $I_D - V_G$ characteristics are shown in Supplementary Fig. 2b. Note that the measurement was performed ~3 months after the initial measurement (Fig. 2d in the main text), and the characteristics have drifted slightly compared to the initial characterization. In particular, the hysteresis voltage had become larger, which indicates an increase in the resistance $R$ of the AlO$_x$ capacitor according to our analysis from the previous section. However, the main features, such as sub-60 mV/decade switching and anti-clockwise hysteresis, are still retained in the new measurement.

From the video, we extract the gate voltage $V_G$ as a function of time $t$ for $V_{DS}$ = 0.4 V, as shown by the red dots in Supplementary Fig. 2c. It is clear that two distinct scan rates of 0.36 V/s ("fast") and 0.208 V/s ("slow") can be extracted from the measured $V_G - t$ curve, which correspond to $I_D$ above and below 1 nA (0.6 nA/μm), respectively. The values for the "fast" and "slow" scans are applied in the simulation for the $V_{int}$ response in the main text. In the forward scan direction (0.5 V to -1 V), sub-60 mV/decade switching is observed when the scan rate transitions from "fast" to "slow" at $I_D$ = 1 nA (green shaded area). In the backward scan (-1 V to 0.5 V), two sub-60 mV/decade switching events are observed, one at $I_D$ = 1 nA and another at $I_D$ = 100 nA, which correspond to the overhead time associated with two measurement range changes, as shown by the orange shaded areas in Supplementary Fig. 2c. Interestingly, the second sub-60 mV/decade event at $I_D$ = 100 nA was not observed in the initial measurement. The cause for the difference is still under investigation, which we speculate might be from some minor changes in the settings of the parameter analyzer in the initial measurement and the new measurement, despite our effort of making them the same.

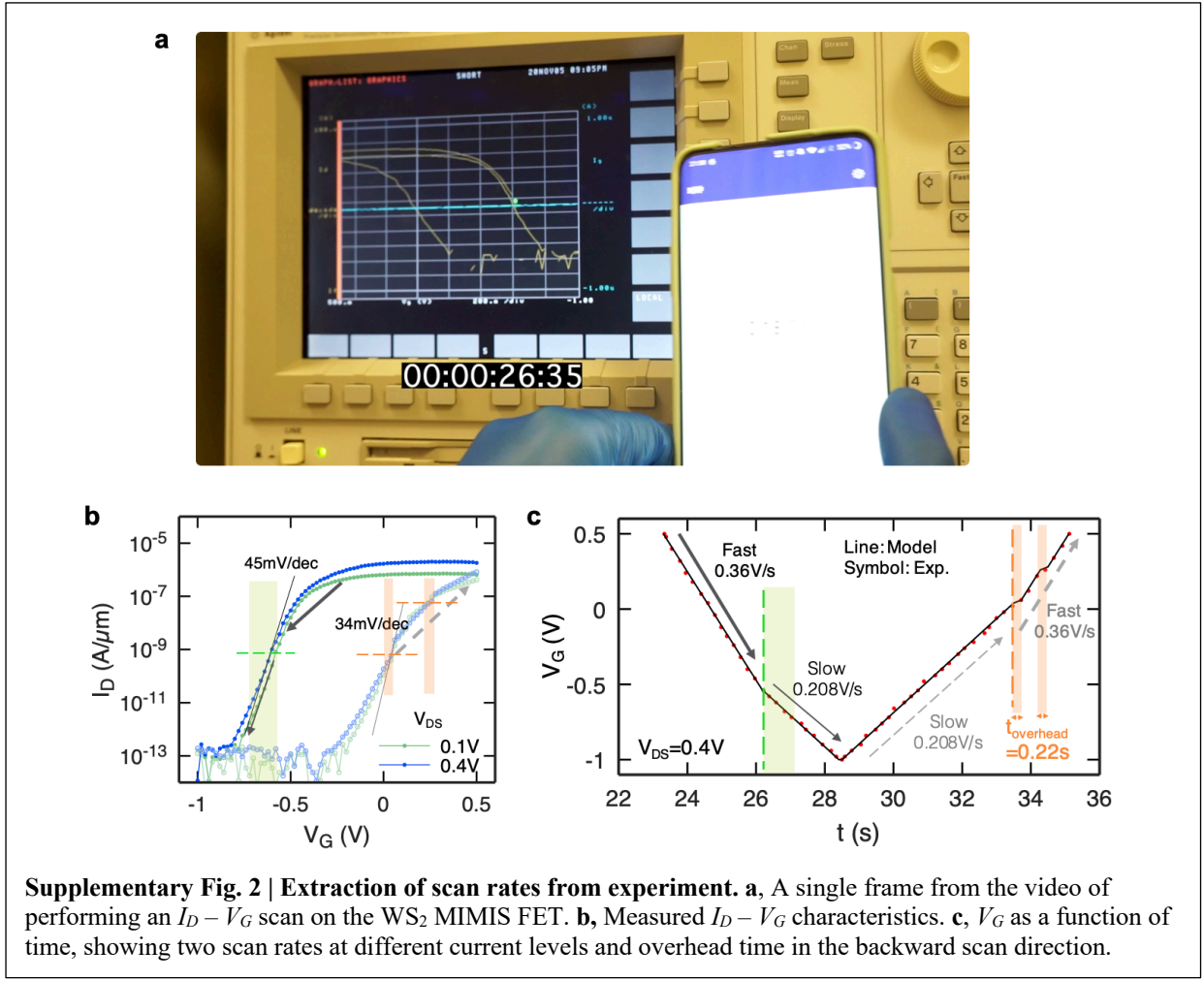

**Supplementary Fig. 2 | Extraction of scan rates from experiment. a**, A single frame from the video of performing an $I_D - V_G$ scan on the WS$_2$ MIMIS FET. **b,** Measured $I_D - V_G$ characteristics. **c**, $V_G$ as a function of time, showing two scan rates at different current levels and overhead time in the backward scan direction.

## 3. Calculation of quantum capacitance $C_q$

Supplementary Fig. 3a shows the schematic of the capacitance model for the WS$_2$ internal FET. The quantum capacitance $C_q$ of 2D WS$_2$ channel is given by:

$$C_q = \frac{\partial Q_{\text{ch}}}{\partial \psi_S} = q^2 \text{DOS}_{2D} f(E_C) A_{\text{device}} = q^2 \frac{g_C m_e^*}{\pi \hbar^2} \frac{1}{1+\exp\left(\frac{E_C - E_F}{k_B T}\right)} A_{\text{device}} \quad (5)$$

Eq. (5) describes the dependence of $C_q$ on the relative position of $E_C$ and Fermi level $E_F$, or equivalently the surface potential $\psi_S$. In order to calculate the $V_{int}$ dependence of $C_q$, we need to determine the relation of $\psi_S$ and $V_{int}$, which is given by:

$$\frac{d\psi_S}{dV_{\text{int}}} = \frac{C_{\text{ox}}}{C_{\text{ox}} + C_q} \quad (6)$$

$$\psi_S = \frac{\Phi_{SB,n} + E_F - E_C}{q} \tag{7}$$

Note that we have chosen flat-band condition as the zero point for the potential $\psi_S$, i.e., $\psi_S = 0$ when $V_{int} = V_{FB}$. For multilayer $WS_2$ in this study, we have:

$$g_C = 6, \; m_e^* = 0.22 \; m_0$$

The only missing parameters are the flat-band voltage $V_{FB}$ and electron Schottky barrier height $\Phi_{SB,n}$. We extract these parameters by fitting experimental $I_D - V_{int}$ data with a Landauer model[1,2], as shown in Supplementary Fig. 3b. Using the Landauer model, we extract the parameters:

$$V_{FB} = -0.46 \; V, \; \Phi_{SB,n} = 0.33 \; eV,$$

and the screening length for Schottky barrier injection $\lambda = 2.6$ nm. Notice that above $V_{th}$ (the extraction of its value will be discussed next), the simulated curve from Landauer model starts to deviate from experimental data. This is because the Landauer approach is based on a ballistic transport picture, while above $V_{th}$, electrons start to populate the channel and a diffusive picture should be adopted for the long channel device ($L \approx 0.6$ μm) in this study. However, the fitting for the data below $V_{th}$, where scattering can be ignored, is already sufficient to extract the critical parameters and we did not try to simulate the on-state of the device quantitatively.

Having determined the values of $V_{FB}$ and $\Phi_{SB,n}$, we calculate $\psi_S$ as a function of $V_{int}$ based on Eqs. (6-7), as shown in Supplementary Fig. 3c. One can see that $\psi_S$ is following $V_{int}$ one-to-one below $V_{th} = -0.26$ V. Above $V_{th}$, channel charges $Q_{ch}$ become significant and $\psi_S$ starts to saturate and deviate from the one-to-one relation.

Finally, we calculate $C_q$ as a function of $\psi_S$ based on Eq. (5), as shown by the blue line in Supplementary Fig. 3d. One would expect $C_q$ to show up as a step function of $\psi_S$ at $T = 0$ K, yet at finite temperatures, the transition becomes more gradual due to thermal broadening. By applying the $\psi_S - V_{int}$ relation shown in Supplementary Fig. 3c, we can determine $C_q$ as a function of $V_{int}$, as shown by the red line in

Supplementary Fig. 3d. Note that $C_{q,max} \gg C_{ox}$, which indicates $C_{MOS}$, the series capacitance of $C_{ox}$ and $C_q$, will quickly saturate to $C_{ox}$ once $V_{int}$ is above $V_{th}$, as shown in Fig. 4b in the main text.

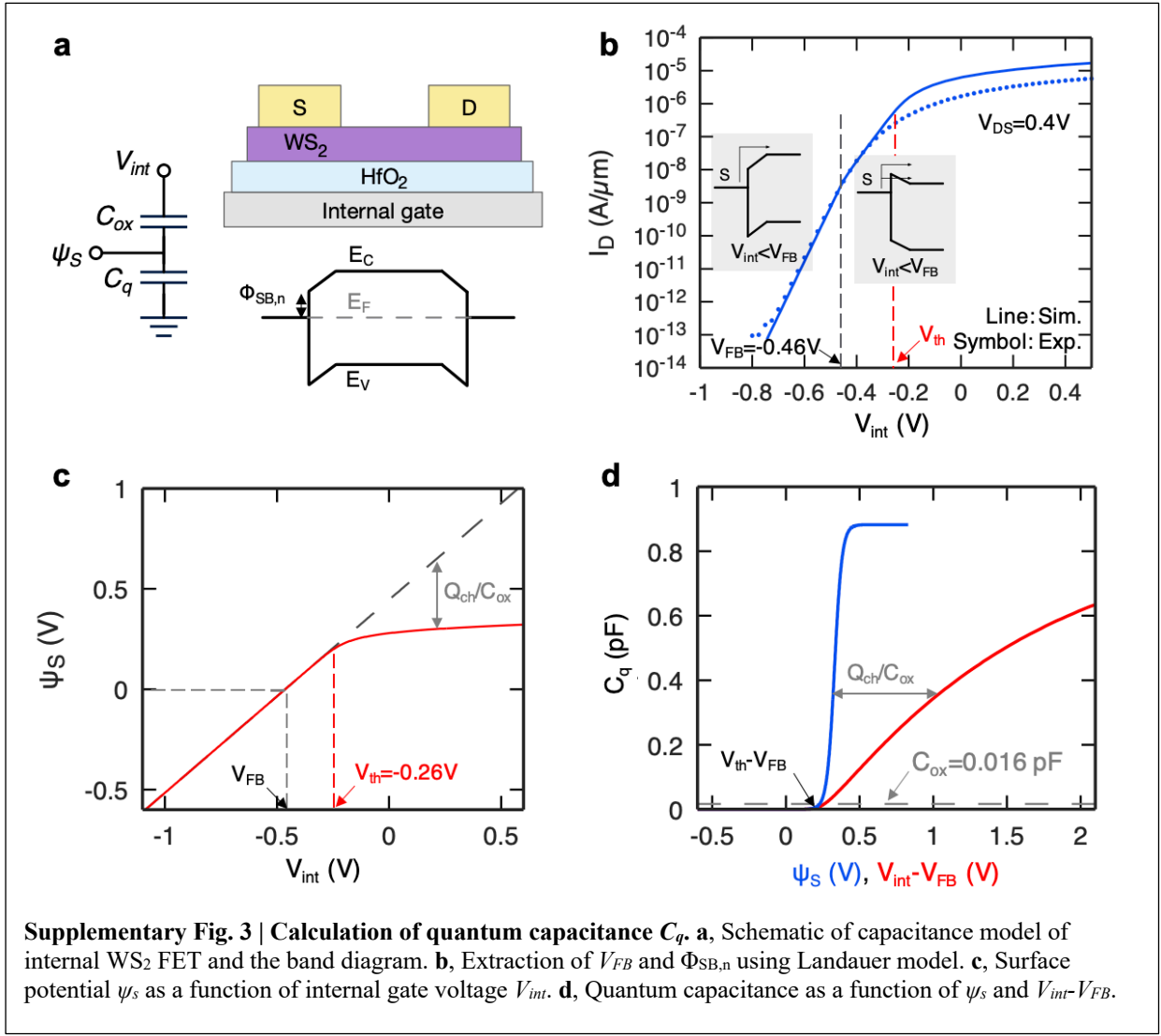

**Supplementary Fig. 3 | Calculation of quantum capacitance $C_q$. a**, Schematic of capacitance model of internal WS$_2$ FET and the band diagram. **b**, Extraction of $V_{FB}$ and $\Phi_{SB,n}$ using Landauer model. **c**, Surface potential $\psi_s$ as a function of internal gate voltage $V_{int}$. **d**, Quantum capacitance as a function of $\psi_s$ and $V_{int}$-$V_{FB}$.